\documentstyle[preprint,aps,epsf,floats,amssymb]{revtex}
\tighten
\widetext

\newcommand{\sect}[1]{\setcounter{equation}{0}\section{#1}}

\begin{document}
\preprint{UPR-0995-T,RUNHETC-2002-13,NI02009-MTH}
{}
\bigskip
\title{{\bf Effective Supergravity for Supergravity Domain Walls}}
{}
\author{
{\bf M. Cveti\v{c}}$^{1,3}$ \thanks{email: cvetic@cvetic.hep.upenn.edu}
and  {\bf N.D. Lambert}$^2$\thanks{email: nlambert@physics.rutgers.edu}}
\medskip
\address{$^1$ Dept. of Physics and Astronomy,
University of Pennsylvania,
Philadelphia,  PA 19104, USA\\
$^2$ Department of Physics and Astronomy,
Rutgers University,
Piscataway, NJ 08854-8019, USA\\
$3$ Isaac Newton Institute for Mathematical Sciences, University of Cambridge,
Cambridge, UK\\
}
\bigskip
 
\maketitle

\begin{abstract}
We discuss the low energy effective action for
the Bosonic and Fermionic zero-modes of a smooth BPS Randall-Sundrum 
domain wall, 
including the induced supergravity on the wall.
The result is a pure supergravity in one lower dimension.
In particular, and in contrast to non-gravitational domain walls or
domain walls in a compact space, 
the zero-modes representing transverse fluctuations of
domain wall have vanishing action.

\end{abstract}

\vfil
\eject

\sect{Introduction}
Supergravity domain walls \cite{Cvetic} have been  recently the
subject of much attention from a variety of different points of view
(for example see 
\cite{Cvetic,BC,ST,Kallosh,GL,LP,CLP,DLS,BCY,Hull,Townsend};  for an earlier
review \cite{CS}).
A central area of study has been that of the so-called Randall-Sundrum
domain walls \cite{RS}, which trap gravity to their worldvolumes. 
It was recognized early on  \cite{BC} that
the tuning used to stabilize such domain walls is simply a supersymmetry
condition. However it has proven very difficult to obtain smooth
four-dimensional 
Randall-Sundrum domain walls from the known supergravities. This
led to several no-go theorems \cite{Kallosh,WZ,GL,MN}.
More recently a five-dimensional supergravity which admits a
smooth Randall-Sundrum domain wall was obtained in \cite{BD}.

In \cite{GL} it was observed that in a 
Randall-Sundrum background the  proposed Goldstone Fermion from broken supersymmetry
diverges on the wall. This result  was interpreted as an explanation
for the absence of smooth Randall-Sundrum domain walls in a large class of
supergravities. However there are  smooth domain walls of that type  in four-dimensional  supergravity \cite{Cvetic}
for which the  the problem described in \cite{GL} 
still occurs, as it also does in the recent five-dimensional example \cite{BD}.
In addition, to the extent of our knowledge, the coupling of 
the zero-modes to the gravity on the wall has not yet been addressed.

In this paper we would like to  extend the discussion of the  Bosonic and
Fermionic zero-modes and obtain the effective action associated with these
modes. In particular, we will reexamine   the dynamics of the zero-modes of
a  supergravity domain wall (of the Randall-Sundrum type). The wall will be 
 viewed as a solitonic object and thus our analysis
will be analogous to the classic  treatment of the low energy motion of  
monopoles in a gauge theory \cite{manton}, which was first applied to
gravity in \cite{GR,FE,Sh}. We will see that in this context  there is no
problem with Fermion zero-modes, indeed these modes do not blow up on the wall.
However, the 
dynamics of Randall-Sundrum type domain walls are qualitatively
different to that of non-gravitational theories, or domain walls in
compact spaces \cite{ADD,DKS,Luty,Bagger}. In particular, 
in contrast to the Higg's mechanism observed in \cite{ADD,DKS} for domain
walls on a circle, we will see that the 
zero-modes that represent the transverse fluctuations are, in a sense,
removed from the
physical spectrum. This is caused by an exact cancellation between the
positive tension of the domain wall and the negative energy density of
bulk anti-de Sitter space. In effect the domain 
wall behaves as if it were tensionless. 

The rest of this paper is organized in the following way. In Section
II  the supersymmetric action (up to bi-linear  Fermionic terms) in $D$
dimensions is given: it contains a gravity  supermultiplet and a matter 
supermultiplet, whose real scalar field  creates a wall. There 
the form of the Fermionic and Bosonic zero-modes is given  and their
effective action, whose prefactors turn out to be zero,  
is discussed.  In Section III   we discuss the properties of the  effective
action  and supersymmetry transformations for the theory reduced on the 
wall.  In Section IV we conclude with  some remarks,  
interpreting the results and suggesting further investigations.

\sect{Domain Walls and Zero-Modes}

For simplicity we assume that we are in $D$-dimensions ($D>3$) with only
gravity, one scalar $\phi$ and their superpartners $\psi_m^i$ and 
$\lambda^i$ active. Here $i$
an internal spinor index which we include for generality. 
We will restrict our attention here to supergravities where the  
supersymmetries takes the form
\begin{eqnarray}
\delta e_{m}^{\ \ \underline n} &=& -\bar\epsilon_i\Gamma^{\underline
  n}\psi_{m}^i + c.c.\ ,\nonumber \\
\delta \phi &=& \bar \epsilon_i\lambda^i+c.c.\ , \nonumber \\
\delta \psi^i_m&=& \nabla_m\epsilon^i+\kappa^{D-2} 
W\Gamma_m\epsilon^i\ \ ,\nonumber\\ 
\delta \lambda^i &=& \left(-{1\over 2}\Gamma^m\partial_m\phi
+(D-2){\partial W\over \partial\phi}\right)\epsilon^i\ ,\nonumber\\ 
\label{susy}
\end{eqnarray}
where 
an underlined index refers to the tangent frame, $m,n=0,1,2,...,D-1$
and $\kappa$ is the $D$-Dimensional Planck length. 
This is certainly not the most general form that one can imagine. 
In particular we have restricted the supersymmetry transformations to
be diagonal in the $i$ indices. However
supergravities of this type have received considerable attention in recent
years and the extension of our analysis to other supergravities is clear.
In particular for the case of $D=5$ a more detailed 
discussion can be found in \cite{Cremmer}.
In even dimensions one expects that there are terms
involving $\Gamma_{D+1}$. However we expect that the analysis presented
here is fairly insensitive to the precise form of the supersymmetry.

An action which is invariant under these supersymmetries, at least to lowest
order in the Fermions, has the form \cite{BCY}
\begin{eqnarray}
S &=&{1\over \kappa^{D-2}}
\int d^{D}x e \left\{R
+\bar\psi_{mi}\Gamma^{mnp}\nabla_n\psi_p^i
- \kappa^{D-2}\partial_m\phi\partial^m\phi
+\kappa^{D-2}\bar\lambda_i \Gamma^m\nabla_m\lambda^i 
-\kappa^{D-2}V(\phi)  \right. \nonumber \\
&&\left.+
2(D-2)\kappa^{D-2} {\partial^2W\over\partial\phi^2}\bar\lambda_i\lambda^i
-(D-2)\kappa^{2D-4} W\bar\lambda_i\lambda^i
- (D-2)W\kappa^{D}\bar\psi_{mi}\Gamma^{mn}\psi_n^i  
\right. \nonumber \\ 
 &&\left.+{1\over2}\kappa^{D}\partial_n\phi(\bar\psi_{mi}\Gamma^n\Gamma^m\lambda^i
+\bar\lambda_i\Gamma^m\Gamma^n\psi_m^i)
+ (D-2)\kappa^{D}{\partial W\over\partial\phi}(\bar\psi_{mi}\Gamma^m\lambda^i
-\bar\lambda_i\Gamma^m\psi_m^i)
+\ldots \right\}\ ,\nonumber\\ 
\label{action}
\end{eqnarray}
where the ellipsis denotes higher order terms in the Fermions. 
The scalar potential is
\begin{equation}
V = 4(D-2)^2\left[\left({\partial W\over\partial\phi}\right)^2-
\kappa^{D-2}\left({D-1\over D-2}\right)W^2\right]\ .
\label{potenial}
\end{equation}
This action reproduces the equations of motion used in \cite{GL}.

We note that the supersymmetric vacuum configurations are simply
$AdS$  spacetimes with
$\phi$ fixed at a critical point  of the superpotential,
$W$, 
so that $V=-4(D-1)(D-2)\kappa^{D-2}W^2\le 0$.
We will use coordinates in which the metric is
\begin{equation}  
ds^2 = dr^2 + e^{2A}\eta_{\mu\nu}dx^\mu dx^\nu\ ,
\label{AdS}
\end{equation}
where $A=-2\kappa^{D-2}W(\phi_0)r$, $r=x^{D-1}$  and  $\mu,\nu=0,...,D-2$. 
The Killing spinors are of the form
\begin{equation}
\epsilon^i_1 = e^{{1\over 2} A}\tilde \eta^i_+\ , \qquad 
\epsilon_2^i = \left(e^{- {1\over 2} A}-2\kappa^{D-2}W(\phi_0)e^{{1\over 2} A}
  x^\mu\Gamma_{\underline\mu} \right)\tilde \eta^i_- \ .
\label{KillingSpinors}
\end{equation}
Here $\tilde \eta^i_\pm$ are constant spinors which satisfy
$\Gamma^r \tilde\eta^i_{\pm}=\pm\tilde\eta^i_\pm$.

Our interest here is in  supersymmetric domain wall solutions which have
the same form for the metric \ref{AdS} but the
scalar field $\phi$ is not constant and $A$ is not a linear function 
\begin{equation}
\phi' = 2 (D-2){\partial W\over \partial\phi}\ ,\qquad
A' =  -2 \kappa^{D-2}W\ ,
\label{wall}
\end{equation}
where  a prime denotes differentiation with respect to $r$. 
This is merely a gravitational version of a BPS kink solution that
interpolates between two supersymmetric vacua. Broken Poincare invariance
implies that it has a Bosonic zero-mode $\tilde r$ corresponding to
the location of the kink, i.e. the general scalar field profile has the form
$\phi(r-\tilde r)$ for any $\tilde r$.  
A particular class of domain walls are the so-called Randall-Sundrum
type \cite{RS}, where $W$ changes sign between the two vacua. These
walls have the interesting feature that  $e^{2A}\sim e^{-4\kappa^{D-2}|Wr|}$ 
as  $r\rightarrow\infty$ leading to localized gravity on the wall. 
For the rest of our discussion  
we will restrict our attention to these types of domain walls.

Next we wish to construct the low energy dynamics associated to the
zero-mode $\tilde r$. From the field theory perspective this is
achieved by allowing $\tilde r$ to depend on the walls' coordinates $x^\mu$.
One then simply evaluates the $D$-dimensional Lagrangian around such a 
background to lowest order in derivatives. 
In a theory with local diffeomorphism invariance one must
be a little more careful. Following \cite{ACGNR} we first note that the
transformation 
\begin{equation}
\delta g_{rr} =0\ ,\quad 
\delta g_{\mu r}=0\ ,\quad 
\delta g_{\mu\nu} = -2A'\tilde r e^{2A}\eta_{\mu\nu}\ ,
\label{diffeo}
\end{equation}
that corresponds to an infinitesimal
but constant shift $r \rightarrow r-\tilde r$ is simply a diffeomorphism.
However if we now let $\tilde r$ depend on $x^\mu$ then \ref{diffeo}
is not a diffeomorphism and therefore we expect it to represent a 
physical mode. 

To continue we consider a variation of the form \ref{diffeo} with an
arbitrary fluctuation $\tilde r(x^\mu)$. To obtain the effective 
Lagrangian we 
substitute  the domain wall solution  back into \ref{action},
and then integrate over $r$. Note that this procedure does not imply
that the full $D$-dimensional equations of motion are satisfied. 
Hence solutions
to the effective action equations of motion do not lift to full solutions
of the $D$-dimensional equations of motion. Instead they represent an
effective description, analogous to the way that 
motion on  monopole moduli space is an effective description of the behaviour
of monopoles. However we must also 
check that the $g_{rr}$ and $g_{\mu r}$ equations of motion are
satisfied identically since the effective action we construct does not
have any fields that represents them (i.e. they act as constraints
on the low energy effective action).

With this prescription we obtain the $(D-1)$-dimensional
Lagrangian (we  postpone including 
variations of the metric on the wall until the next section)
\begin{equation}
{\tilde {\cal L}}_B= {\tilde {\cal L}}_W-4(D-2)^2\int dr e^{(D-3)A}
\left(\left({\partial W\over\partial\phi}\right)^2-\kappa^{D-2}
\left({D-3\over D-2}\right)W^2\right)
\eta^{\mu\nu}\partial_\mu \tilde r\partial_\nu \tilde r\ ,
\label{Bact}
\end{equation}
where ${\tilde {\cal L}}_W$ is the integral over $r$ of the Lagrangian
evaluated on the wall solution
\begin{equation}
{\tilde {\cal L}}_W= -8(D-2)^2\int dr e^{(D-1)A}
\left( \left({\partial W\over \partial\phi}\right)^2
-\kappa^{D-2}\left({D-1\over D-2}\right)W^2\right)
\ .
\label{LW}
\end{equation}
However we note that, for any $d$, 
\begin{equation}
2(D-2)\left(\left({\partial W\over \partial\phi}\right)^2
-\kappa^{D-2}\left({d\over D-2}\right)W^2 \right)e^{dA}= 
{d\over dr}\left(We^{dA}\right)\ .
\label{opps}
\end{equation}
Hence for a Randall-Sundrum domain wall, 
due to the exponential fall-off of the metric at large $r$, 
${\tilde {\cal L}}_B={\tilde {\cal L}}_W=0$.  Thus the effective
action for fluctuations of the wall vanishes, even though these
fluctuations are not diffeomorphisms. Note that this integral vanishes 
only if $\kappa \ne 0$.

Now we wish to consider the Fermionic properties of a domain wall.
One can readily check that $\epsilon^i_1$ is still a Killing spinor of
the domain wall background,
whereas  $\epsilon^i_2$ is not. 
The fact that half of the supersymmetries of
the $AdS$ vacuum are preserved by the wall implies that the
Bosonic zero-mode has a superpartner, so that the preserved
supersymmetry is linearly realized.
In particular the broken supersymmetry 
creates this
Fermionic zero-mode. However
in supergravity, or any theory with 
local symmetry, there
are an infinite number of broken supersymmetries. Any such spinor
is a linear combination
\begin{equation}
\epsilon^i  = F_+\tilde \eta^i_+ + F_-\tilde \eta^i_-\ ,
\label{lincom} 
\end{equation}
with arbitrary
coefficients $F_+(x^m)$ and $F_-(x^m)$.
To continue then let us outline two natural choices for the physical
Goldstino mode.

At first thought we should choose the resulting Goldstino to 
respect the same symmetries as the wall, i.e. 
$\partial_\mu=0,\ \psi^i_\mu=0$. 
This is obtained by acting with supersymmetry that is preserved by 
AdS space but broken by the wall: 
$F_+=0$,  
$F_- = \kappa^{1-D}W^{-1}(e^{- {1\over 2} A}-2\kappa^{D-2}We^{{1\over 2} A}
  x^\mu\Gamma_{\underline\mu})$
and yields
\begin{equation}
\lambda^i = {2(D-2)\over \kappa^{D-1}}{1\over W}{\partial W\over \partial\phi}
e^{-{1\over2}A}\tilde \eta^i_-\ ,\qquad
\psi^i_r =-{2(D-2)\over \kappa^{D-1}}\left( {1\over W}{\partial W\over \partial\phi}\right)^2
e^{-{1\over2}A}\tilde\eta^i_-\ ,\qquad \psi^i_\mu=0\ .
\label{one}
\end{equation}
This is precisely the Goldstino found in \cite{GL} and diverges if
the superpotential $W$ changes sign. 
This will mean that when we construct the effective action for the
Fermionic zero-mode, found by letting $\tilde\eta^i_-$ become a
field which depends on  $x^\mu$, we find
\begin{equation}
{\tilde {\cal L}}_F= 4(D-2)^2\kappa^{2-2D}\int dr e^{(D-3)A}
\left({1\over W}{\partial W\over\partial\phi}\right)^2
\bar{\tilde\eta}_{-i}
\Gamma^{\underline \mu}\partial_\mu \tilde\eta^i_-  +\ldots\ ,
\label{actone}
\end{equation}
where the ellipsis denotes cross terms involving
$\bar{\tilde\eta}_{-i}\partial_{\nu}\tilde r\eta_-^i$.
The existence of these terms indicates that a field redefinition is
needed to put the action into a standard form.
Thus in a Randall-Sundrum type of domain wall, where $W$ passes through zero, 
the kinetic term for such a  Fermionic zero-modes diverges.
This seems contradictory since the kinetic term for the Bosonic
zero-mode $\tilde r$ is well behaved, indeed it vanishes. 
In \cite{GL} this was used as
an indication that $W$ cannot change sign in a well-defined
supergravity. However it is clear that the divergence is caused by
the choice $\epsilon^i = W^{-1}\epsilon^i_2$. 

A better choice is to find a Goldstino mode with $\partial_\mu=0$
but $\psi^i_\mu\ne 0$. This can be done by simply acting on the
domain wall with the supersymmetry generated by
$e^{-{1\over2}A}\tilde\eta^i_-$ and yields
\begin{equation}
\lambda^i = 2(D-2){\partial W\over \partial\phi}
e^{-{1\over2}A}\tilde\eta_-^i\ ,\qquad
\psi^i_r=0\ ,\qquad
\psi^i_\mu =2\kappa^{D-2} W
\Gamma_{\underline \mu}e^{{1\over2}A}\tilde\eta^i_-\ .
\label{three}
\end{equation}
This solution is much nicer.  Indeed if we evaluate the effective 
action for it we find
\begin{equation}
{\tilde {\cal L}}_F= 4(D-2)^2\int dr e^{(D-3)A}
\left(\left({\partial W\over\partial\phi}\right)^2-
\kappa^{D-2}\left({D-3\over D-2}\right)W^2\right)
\bar{\tilde\eta}_{-i}\Gamma^{\underline \mu}\partial_\mu \tilde\eta^i_- \ .
\label{actthree}
\end{equation}
Thus we encounter precisely the same, vanishing,  integral that we
obtained for the Bosonic zero-mode.
In addition the cross terms involving 
$\bar{\tilde\eta}_{-i}\partial_{\nu}\tilde r\tilde\eta^i_-$ 
are total derivatives and can be 
discarded. Therefore \ref{three} seems  
to be the correct choice of
Goldstino that is linearly related to $\tilde r$.

It is instructive to contrast this discussion with the case of
a domain wall a in non-gravitating theory.
Specifically,  we take the  flat space limit $\kappa =0$ and set 
$e_m^{\ \ \underline n}=\delta_m^{\ \ \underline n}$,  $\psi_m^i=0$.
Thus the original $D$-dimensional action \ref{action} simplifies to
\begin{eqnarray}
S_{\kappa=0} &=&
-\int d^{D}x  \left\{
\partial_m\phi\partial^m\phi 
+ 4(D-2)^2\left({\partial W\over \partial \phi}\right)^2
-\bar\lambda_i \Gamma^m\partial_m\lambda^i -
2(D-2) {\partial^2W\over\partial\phi^2}\bar\lambda_i\lambda^i\right\}
\label{faction}
\end{eqnarray}
The supersymmetry transformations are easy determined from \ref{susy}
by setting $\kappa=0$ and they reduce to the rigid supersymmetry
$\delta \tilde\eta^i_- = {1\over2} \tilde\Gamma^\mu\partial_\mu \tilde r$
, $\delta \tilde r = -\bar{\tilde\epsilon}_{+i}\tilde\eta_-^i 
+c.c.$.
However, integrals of terms of the form \ref{opps} no longer vanish 
and we instead find
the effective action of a free scalar $\tilde r$ and its superpartner 
$\tilde\eta^i_-$
\begin{equation}
\tilde{\cal L}_{\kappa=0}  
= -8(D-2)^2\int dr \left({\partial W\over\partial\phi}\right)^2
\left(1+{1\over 2}\partial_{\mu} \tilde r\partial^{\mu} \tilde r\ 
-{1\over 2}\bar{\tilde\eta}_{-i}\tilde\Gamma^{\mu}\partial_{\mu} \tilde\eta^i_-\right)\ .
\label{flatact}
\end{equation}
The integral over $r$ is again a total derivative and can be evaluated to be  
$4(D-2)|W(r=\infty)-W(r=-\infty)|$, which is simply the tension of the 
domain wall.  Thus the low energy dynamics of
the zero-modes in the supergravity domain wall can not be
continuously reduced to the flat space limit by making $\kappa$
arbitrarily small. 

\sect{Supergravity on the Wall}

In this section we are interested in understanding how the zero-modes
found above, which describe the fluctuations of the wall,  couple to the 
gravitational fields. We will be primarily interested in whether or
not the full effective action of the wall can be identified with a 
$(D-1)$-dimensional supergravity.
Previous studies have discussed the reduction of the supergravity to
the wall \cite{LP,DLS,CLP,DLS} 
however these have not included the zero-modes, i.e. they treat the
wall as rigid. We also seek to further 
justify the choice \ref{three} as the correct Fermionic zero-mode.

More precisely we wish to reduce the $D$-dimensional supersymmetry involving
$e_m^{\ \underline n},\phi,\lambda^i,\psi^i_m$ to $(D-1)$-dimensional
supersymmetry involving $\tilde r,\tilde \eta^i_-$ 
and the bulk supergravity fields,
suitably dimensionally reduced. In particular we consider the following 
standard ansatz for the Bosonic fields
\begin{eqnarray}
\phi &=& \phi(r-\tilde r)\ , \nonumber\\
A&=&A(r-\tilde r)\ ,\nonumber\\
e_m^{\ \ \underline n} &=& \left(
\matrix{1&e^{\alpha A}\tilde A_{\mu}\cr
0&e^A\tilde e_{\mu}^{\ \underline \nu}\cr }\right)\ , \nonumber\\
\label{ansatzB}
\end{eqnarray}
where $\phi$ and $A$ continue to satisfy the domain wall Bogomoln'yi
equations \ref{wall} and $\alpha$ is a parameter that we will fix shortly. 
In what follows we use tildes to denote $(D-1)$-dimensional fields and
$\tilde \Gamma_{\mu}=\tilde e_{\mu}^{\ \underline \nu}
\Gamma_{\underline \nu}$.
Our next task is to determine the correct form for the Fermions
so as to obtain a supersymmetry
acting only on the $(D-1)$-dimensional fields living on the wall.

First we note that the action \ref{action}, 
when expressed in terms of 
the new fields, is still invariant under the supersymmetry 
\ref{susy}. 
Although the transformations given in \ref{susy} are symmetries for
any $\epsilon^i$, we are only interested in those 
that preserve the domain wall. Thus we are led to introduce the $(D-1)$-dimensional
supersymmetry generator $\tilde \epsilon^i_+$ as
\begin{equation}
\epsilon^i =  e^{{1\over 2}A}\tilde\epsilon^i_+  \ ,
\label{susytwo} 
\end{equation}
where $\Gamma_r\tilde\epsilon^i_+ =
\tilde\epsilon^i_+$.

With the form for $\lambda^i$ given in \ref{three} 
the $\delta\phi$ and $\delta \lambda^i$ variations
simply reduce to
\begin{eqnarray}
\delta \tilde r &=& -\bar{\tilde\epsilon}_{+i}\tilde\eta_-^i 
+c.c.\ ,\nonumber\\
\delta \tilde\eta^i_- &=& {1\over2} \tilde\Gamma^\mu\left(
\partial_\mu \tilde r+e^{\alpha A}\tilde A_\mu\right)
\tilde\epsilon^i_+\ .
\nonumber\\
\label{zmsusy}
\end{eqnarray}
Recall that we restrict to terms that are at most quadratic in the Fermi
fields. 
The supersymmetry reduces
to an expression that involves only the $(D-1)$-dimensional fields only
if $\alpha =0$. 
That this  occurs at all is due to 
the precise form for the $\lambda^i$ Goldstino
in \ref{three} and does not occur if we used 
any other form. The appearance of $\tilde A_\mu$ is reminiscent of a  
Higg's mechanism  where $A_\mu$ ``eats'' 
$\partial_\mu\tilde r$. Thus we  
see that $\tilde r$ and $\tilde\eta^i_-$ defined by \ref{three} 
are indeed superpartners on the wall.

Next we must ensure that the  
gauge choice $e_r^{\ \underline r}=1$ and $e_{\mu}^{\ \underline r}=0$
is preserved by the supersymmetries generated by $\tilde\epsilon^i_+$.
This implies that $\psi^i_r=0$. 
However to preserve $\psi^i_r=0$ we must have
\begin{equation}
0=\delta \psi^i_{r}
= -{1\over 8}\tilde F^{\mu\nu}\tilde\Gamma_{\mu\nu}e^{-{3\over 2}A}
\tilde\epsilon_+^i\ ,
\label{psir}
\end{equation} 
where here, and in what follows, we have set $\alpha=0$.
Thus to preserve supersymmetry on the wall we must set $\tilde F_{\mu\nu}=0$.
Note that this does not necessarily imply  $\tilde A_\mu=0$.
Another way to see this restriction arises in the physically interesting
cases of $D \le 5$. It is not hard to see that the reduction of the 
Einstein-Hilbert action in $D$ dimensions using the ansatz \ref{ansatzB} leads
to the kinetic term 
\begin{equation}
-{1\over 4 \kappa^{D-2}}\int dr e^{(D-5)A}\tilde F^2\ ,
\label{Fkinetic}
\end{equation}
for the graviphoton. Clearly in $D \le 5$ the integral over $r$ is
infinite. From the point of view of 
the theory on the wall this can be interpreted as saying that
the $(D-1)$-dimensional electromagnetic coupling constant vanishes and hence 
Maxwell's equation is simply $\tilde F_{\mu\nu}=0$.

Next we consider the gravitini $\psi_\mu^i$. In particular, using \ref{three}
as a guide,  we
let $\psi_\mu^i=2\kappa^{D-2}We^{{1\over 2}A}\tilde\Gamma_{\mu}\tilde\eta^i_- +
+e^{{1\over 2} A}\tilde\psi^i_{\mu +} +e^{\beta A}\tilde\psi^i_{\mu-} 
+e^{\gamma A}\tilde\Gamma_\mu\tilde\chi^i_+$, where $\beta$ and $\gamma$ are 
to be  determined. Note that $\tilde\chi^i_+$ and $\tilde\eta_-^i$ can be 
distinguished from each other by their chiralities and the coefficient
of $\tilde\psi^i_{\mu +}$ will be justified by the calculations which follow.
Considering the variation of $\psi_\mu^i$ we find
\begin{equation}
\delta \tilde\psi^i_{\mu_+} 
+ e^{(\beta-{1\over2})A}\delta \tilde\psi^i_{\mu -} 
= \tilde\nabla_\mu\tilde\epsilon^i_+\ ,\quad
\delta\tilde\chi^i_+=0\ ,
\label{deltapsi}
\end{equation}
where we have used the restriction $\tilde F_{\mu\nu}=0$.
Next we substitute this ansatz into the variation of 
$e_{\mu}^{\ \ \underline r}=\tilde A_\mu$ to find
\begin{equation}
\delta \tilde A_\mu = 
e^{({1\over2}+\beta)A}\bar{\tilde \epsilon}_{+i} \tilde\psi^i_{\mu-}
+e^{({1\over2}+\gamma)A}\bar{\tilde \epsilon}_{+i} 
\tilde\Gamma_\mu\tilde\chi^i_++c.c.\ .
\label{deltaA}
\end{equation}
To ensure that the right hand side is independent of $r$ we must
set $\beta=\gamma=-1/2$. 
Thus $\tilde \chi^i_+$ plays the role of a graviphotini
since $\delta \tilde A_\mu=\bar{\tilde\epsilon}_{+i}
\tilde\Gamma_\mu\tilde\chi^i_+$ while $\tilde \psi^i_{\mu -}$
is sterile in the sense that \ref{deltapsi} 
implies  $\delta\tilde\psi^i_{\mu -}=0$. 
However, as with $\tilde A_{\mu}$ above, 
it is easy to see that the $\tilde\psi^i_{\mu -}$ and 
$\tilde\chi^i_+$ kinetic terms in the
effective action are infinite if $D\le 5$ and hence they must be set to zero.
This also follows by supersymmetry for any $D$ 
since we must have $\tilde F_{\mu\nu}=0$ for all
variations $\tilde\epsilon^i_+$. Thus we
set $\tilde\psi^i_{\mu -}=\tilde\chi^i_+=0$.

Lastly we can obtain the variation of the vielbein on the wall
\begin{equation}
\delta {\tilde e}_\mu^{\ \underline \nu} 
= -\bar{\tilde\epsilon}_{+i}\tilde 
\Gamma^{\underline \nu}\tilde\psi^i_{\mu +} 
-2 \kappa^{D-2}W\bar{\tilde\epsilon}_{+i}\tilde 
\Gamma^{\underline {\nu\lambda}}\tilde\eta^i_-\tilde e_{\mu\underline\lambda}
 + c.c.\ .
\label{esusy}
\end{equation}
Here we find that the variation 
involves the $r$-dependent term 
$2\kappa^{D-2}W\bar{\tilde\epsilon}_{+i}\tilde 
\Gamma^{\underline {\nu\lambda}}\tilde\eta_-^i\tilde 
e_{\mu\underline\lambda}+c.c.$. 
To continue let us first 
summarize our calculations so far.  We have found the following
ansatz for the Fermions
\begin{eqnarray}
\lambda^i &=& 2(D-2){\partial W\over \partial\phi}
e^{-{1\over2}A}\tilde\eta_-^i\ ,\nonumber\\
\psi^i_\mu &=& 2\kappa^{D-2}We^{{1\over 2}A}\tilde\Gamma_{\mu}\tilde\eta^i_- +
e^{{1\over 2}A}\tilde\psi^i_{\mu+}\ ,\nonumber\\
\psi^i_r &=& 0\ ,\nonumber\\
\label{ansatzF}
\end{eqnarray}
which leads to the symmetry 
\begin{eqnarray}
\delta  {\tilde e}_\mu^{\ \underline \nu} 
&=& -\bar{\tilde\epsilon}_{+i}\tilde 
\Gamma^{\underline \nu}\tilde\psi^i_{\mu +} 
-2\kappa^{D-2}W\bar{\tilde\epsilon}_{+i}\tilde 
\Gamma^{\underline {\nu\lambda}}\tilde\eta^i_-\tilde e_{\mu\underline\lambda}
 + c.c.\  \ ,\nonumber\\
\delta \tilde r &=& -\bar{\tilde\epsilon}_{+i}\tilde\eta_-^i 
+c.c.\ ,\nonumber\\
\delta \tilde A_\mu &=& 0
\ ,\nonumber\\
\delta \tilde\eta^i_- &=& {1\over2} \tilde\Gamma^\mu\left(
\partial_\mu \tilde r+\tilde A_\mu\right)
\tilde\epsilon^i_+\ , \nonumber\\
\delta \tilde\psi^i_{\mu+} &=& \tilde\nabla_\mu\tilde\epsilon^i_+\ ,\nonumber\\
\label{wallsusy}
\end{eqnarray}
where we have imposed $\tilde F_{\mu\nu}=0$.

The additional term in the vielbein variation is puzzling.
However it is in fact rather harmless. To see this we may obtain the
effective action for the $(D-1)$-dimensional fields by substituting
in the ansatz \ref{ansatzB},\ref{ansatzB} into \ref{action}. 
For a Randall-Sundrum domain wall this yields 
\begin{equation} 
\tilde{\cal L} = {1\over  \kappa^{D-2}}
\int dr\ e^{(D-3)A} \tilde e \left\{
\tilde R 
+ \bar{\tilde\psi}_{\mu + i}\tilde\Gamma^{\mu\nu\rho}
\tilde\nabla_{\nu}\tilde{\psi}^i_{\rho +}\right\} \ .
\label{fact}
\end{equation}
Note that in deriving \ref{fact} we have discarded several 
terms whose integral
over $r$ vanishes due to the asymptotic fall off of the metric.
We have also 
checked that the $g_{rr}$ and $g_{\mu r}$ equations of motion are
satisfied identically.  

The action \ref{fact} 
is the minimal supergravity Lagrangian in $(D-1)$ dimensions.
The integral over $r$ in \ref{fact} is finite and
is simply absorbed into Planck's constant in $(D-1)$-dimensions.
In addition $\tilde r$, $\tilde A_\mu$ and $\tilde\eta^i_-$ have all
disappeared from the action, i.e. they do not represent any physical
modes of the low energy effective dynamics.
Therefore without loss of generality we may set  
$\tilde A_\nu=-\partial_{\nu}\tilde r$ and 
$\tilde \eta_-^i=0$.
In this case the 
final supersymmetry on the wall is just that of a minimal supergravity
\begin{eqnarray}
\delta  {\tilde e}_\mu^{\ \underline \nu} 
&=& -\bar{\tilde\epsilon}_{+i}\tilde 
\Gamma^{\underline \nu}\tilde\psi^i_{\mu +} 
 + c.c.\  \ ,\nonumber\\
\delta \tilde\psi^i_{\mu+} &=& \tilde\nabla_\mu\tilde\epsilon^i_+\ .\nonumber\\
\label{Fwallsusy}
\end{eqnarray}
We note that for an arbitrary choice of $\tilde r(x)$
and $\delta g_{\mu r}=\tilde A_\mu$ the corresponding variation 
\ref{diffeo}  of the wall is 
not a diffeomorphism. However in the special case that  
$\tilde A_\mu =-\partial_\mu\tilde r$, 
where the supersymmetry of transformation rules are simple,  
then the variation is a diffeomorphism. 

In a sense the Bosonic zero-mode is eaten by the graviphoton, in a 
manner similar to \cite{ADD,DKS} and similarly the gravitini has, in 
a sense, eaten the Fermionic zero-mode. 
However  the graviphoton and (part of the) gravitini are not
really massive but rather have been frozen out all together.
Furthermore, and in contrast to the Higg's mechanism, the
kinetic terms for $\tilde r$ and $\tilde \eta_-^i$ have disappeared.
Thus, rather than becoming components of some massive fields, the zero-modes
are completely removed from the physical spectrum.
We can understand
the vanishing of the kinetic terms for $\tilde r$ and $\tilde \eta_-^i$
as an exact cancellation between
the positive tension of the domain wall, given by the first term
on the left hand side of \ref{opps}, and the negative energy density
of anti-de Sitter space, given by the second term on the
left hand side of \ref{opps}.

We would also like to contrast our results to those  
obtained in \cite{Luty,Bagger} where
the effective action for (infinitely thin) Randall Sundrum  domain walls,
where the transverse  dimension was a compact $Z_2$ orbifold. In this case
prefactor for the  a kinetic  energy term of the modulus
is non-zero. Note however, that in this case the transverse direction is
finite (due to the orbifold compactification) whereas here the vanishing of
the kinetic terms arises because of the infinite extra dimension.
In addition in this case,  the choice of 
the boundary conditions seems to remove the the Bosonic
zero mode $\tilde r$. Consequently the 
authors of \cite{Luty,Bagger} allowed for a more
general fluctuation of the metric component $g_{r\,r}$ and this may be an
interesting direction to further explore within our context.
We  also note that the action that we have
considered should be viewed as a truncation of supergravity to the
sector most relevant to the domain wall. In general we expect that there
will be other scalars and $p$-form fields which will contribute to
the low energy dynamics. However we do not expect that these fields 
will affect the dynamics of $\tilde r$ that we discussed here.

\section{Conclusions}

In this paper we have evaluated the effective action for the Bosonic
and Fermionic zero-modes of a Randall-Sundrum domain wall. Our result
was  simply \ref{fact}, i.e. pure supergravity. In particular the zero-mode,
and its superpartner, 
that are normally identified with transverse fluctuations of the wall
were found to have a vanishing action.

Given this somewhat surprising result it seems appropriate to mention 
some aspects of the approximation
that we have used to obtain the effective action. In particular we
considered slow motion on the moduli space
by allowing the domain wall to fluctuate  and then
evaluating the action to second order in derivatives. As a consequence 
some modes, such as the graviphoton, are frozen, i.e.  
their kinetic terms diverge. This is a familiar
effect in soliton dynamics and implies that the low energy dynamics
is restricted to a subspace of the full moduli space. In particular
the freezing of half of the gravitini leads to
the reduced supersymmetry in the effective theory.
Other modes, such as
the fluctuations of gravity along the wall, are described by the appropriate
action in the lower dimension and lead to non-trivial dynamics at
low energy.
However the  transverse fluctuations have not been frozen out, nor do they
have a kinetic term.
Instead they disappear from the action regardless of the form they take.
In effect, even though these modes are not diffeomorphisms in the 
full theory, they behave like diffeomorphisms from the perspective of 
the effective theory. 

Note that the effective action does not contain the full dynamics
of the theory, just those that are relevant to small fluctuations
about the soliton.  It would be interesting
to further understand the validity of this approximation and
determine if there are applications to brane-world scenarios 
where the low energy dynamics of our world are insensitive to the presence
of the extra-dimension through any of the fluctuations of the domain wall.

\bigskip
{\noindent \bf Acknowledgments}\\

We would like to thank N. Arkani-Hamed, H. Liu, R. Myers, J-L. Lehners,
D. Tong   and particularly  K. Stelle for discussions. N.D.L. Would like to
thank the University of Pennsylvania for its hospitality where this work
was initiated and was also partially supported by a PPARC Advanced Fellowship
at the Department of Mathematics, King's College London and the 
Isaac Newton Institute of Cambridge University. M.C. is supported by
the DOE grant  EY-76-02-3071 and by the UPenn Class of 1965 Endowed Term Chair
and would like to thank High Energy Theory group of Rutgers University,  
and the Isaac Newton Institute of Cambridge University for
hospitality and support during the course of the work.


\bigskip




\begin{thebibliography}{10}

\bibitem{Cvetic}
M.~Cveti\v c, S.~Griffies and S.~J.~Rey,
{\it Static domain walls in N=1 supergravity},
Nucl.\ Phys.\ B {\bf 381}, 301 (1992), 
hep-th/9201007.

\bibitem{BC}
K.~Behrndt and M.~Cveti\v c,
{\it Supersymmetric domain wall world from D = 5 simple gauged supergravity},
Phys.\ Lett.\ B {\bf 475}, 253 (2000), hep-th/9909058.

\bibitem{ST}
K.~Skenderis and P.~K.~Townsend,
{\it Gravitational stability and renormalization-group flow}, 
Phys.\ Lett.\ B {\bf 468}, 46 (1999), hep-th/9909070.

\bibitem{Kallosh}
R.~Kallosh and A.~D.~Linde,
{\it Supersymmetry and the brane world},
JHEP {\bf 0002}, 005 (2000), hep-th/0001071.

\bibitem{GL}G.W. Gibbons and N.D. Lambert, {\it Solitons and Domain
    Walls in Odd Dimensions}, Phys. Lett. {\bf B488} (2000) 90, 
hep-th/0003197.


\bibitem{LP}
H.~L\"u and C.~N.~Pope,
{\it Branes on the brane},
Nucl.\ Phys.\ B {\bf 598}, 492 (2001), hep-th/0008050.

\bibitem{CLP}
M.~Cveti\v c, H.~L\"u and C.~N.~Pope,
{\it Brane-world Kaluza-Klein reductions and branes on the brane}, 
hep-th/0009183.

\bibitem{DLS}
M.~J.~Duff, J.~T.~Liu and W.~A.~Sabra,
{\it Localization of supergravity on the brane},
Nucl.\ Phys.\ B {\bf 605}, 234 (2001),
hep-th/0009212.

\bibitem{BCY}F. Brito, M. Cveti\v c and S-C. Yoon,
{\it From Thick to Thin Domain Walls in Supergravity}, 
Phys.\ Rev.\ D {\bf 64}, 064021 (2001), hep-ph/0105010.

\bibitem{Hull}
C.M. Hull, {\it Domain Wall and de Sitter Solutions of Gauged 
Supergravity}, 
JHEP {\bf 0111}, 061 (2001), hep-th/0110048.


\bibitem{Townsend}
P.K. Townsend, {\it Quintessence from M-theory}, 
JHEP {\bf 0111}, 042 (2001),
hep-th/0110072.

\bibitem{CS}
M.~Cveti\v c and H.~H.~Soleng,
{\it Supergravity domain walls},,
Phys.\ Rept.\  {\bf 282}, 159 (1997), hep-th/9604090.

\bibitem{RS} 
L. Randall and R. Sundrum, 
{\it An Alternative to 
Compactification}, Phys. Rev. Lett. {\bf  83} (1999) 4690, hep-th/9906064.

\bibitem{WZ}
M.~Wijnholt and S.~Zhukov,
{\it On the uniqueness of black hole attractors}, hep-th/9912002.

\bibitem{MN}
J.~Maldacena and C.~Nunez,
{\it Supergravity description of field theories on curved manifolds and a no  go theorem},
Int.\ J.\ Mod.\ Phys.\ A {\bf 16}, 822 (2001), 
hep-th/0007018.

\bibitem{BD}
K.~Behrndt and G.~Dall'Agata,
{\it Vacua of N = 2 gauged supergravity derived from non-homogeneous  quaternionic spaces}, hep-th/0112136.


\bibitem{manton}
N.~S.~Manton,
{\it A Remark On The Scattering Of Bps Monopoles},
Phys.\ Lett.\ B {\bf 110}, 54 (1982).

\bibitem{GR}
G.~W.~Gibbons and P.~J.~Ruback,
{\it The Motion Of Extreme Reissner-Nordstrom Black Holes In The Low Velocity Limit},
Phys.\ Rev.\ Lett.\  {\bf 57}, 1492 (1986).

\bibitem{FE}
R.~C.~Ferrell and D.~M.~Eardley,
{\it Slow Motion Scattering And Coalescence Of Maximally Charged Black Holes},
Phys.\ Rev.\ Lett.\  {\bf 59}, 1617 (1987).

\bibitem{Sh}
K.~Shiraishi,
{\it Moduli space metric for maximally charged dilaton black holes},
Nucl.\ Phys.\ B {\bf 402}, 399 (1993).

\bibitem{ADD} N. Arkani-Hamed, S. Dimopoulos and G. Dvali, 
{\it The Hierarchy Problem and New Dimensions at a Millimeter}, 
Phys. Lett. {\bf B429} (1999) 263, hep-th/9803315.

\bibitem{DKS} G. Dvali, I.I. Kogan and M. Shiftman, 
{\it Topological Effects in 
Our Brane World From Extra Dimensions}, Phys. Rev. 
{\it D62} (2000) 106001, hep-th/0006213.


\bibitem{Luty}
M.~A.~Luty and R.~Sundrum,
{\it Hierarchy stabilization in warped supersymmetry},
Phys.\ Rev.\ D {\bf 64}, 065012 (2001), hep-th/0012158.


\bibitem{Bagger}
J.~Bagger, D.~Nemeschansky and R.~J.~Zhang,
{\it Supersymmetric radion in the Randall-Sundrum scenario,}
JHEP {\bf 0108}, 057 (2001), hep-th/0012163.

\bibitem{ACGNR}
T.~Adawi, M.~Cederwall, U.~Gran, B.~E.~Nilsson and B.~Razaznejad,
{\it Goldstone tensor modes},
JHEP {\bf 9902}, 001 (1999), hep-th/9811145.

\bibitem{Cremmer}
E.~Cremmer,
{\it Supergravities In 5 Dimensions}, 
in {\it C80-06-22.1.1}
LPTENS 80/17
{\it Invited paper at the Nuffield Gravity Workshop, 
Cambridge, Eng., Jun 22 - Jul 12, 1980}.

\end{thebibliography}
\end{document}